\documentclass[preprint,aps,prd,12pt,preprintnumbers,eqsecnum,nofootinbib]{revtex4-2}
\usepackage{xcolor}
\usepackage{braket}
\usepackage{amssymb,amsmath,amsfonts,comment,latexsym,graphicx,epstopdf,float}
\usepackage{color}
\usepackage{fancyvrb}
\usepackage{chngcntr}
\usepackage{etoolbox}
\usepackage{ulem}
\usepackage{array}
\usepackage{bbold}
\usepackage[labelformat=empty]{subfig}
\usepackage{slashed}
\usepackage{multirow}

\usepackage[colorlinks=true,citecolor=blue,linkcolor=red,filecolor=cyan,urlcolor=magenta]{hyperref}

\usepackage{float}

\newcommand{\be}{\begin{equation}}
\newcommand{\ee}{\end{equation}}
\newcommand{\ba}{\begin{eqnarray}}
\newcommand{\ea}{\end{eqnarray}}

\newcommand{\grts}{\raise.3ex\hbox{$>$\kern-.75em\lower1ex\hbox{$\sim$}}}
\newcommand{\lets}{\raise.3ex\hbox{$<$\kern-.75em\lower1ex\hbox{$\sim$}}}

\newcommand{\OneGluonVertex}{ \makebox{\raisebox{-1.6cm}{\includegraphics{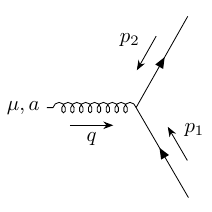}~}} }
\newcommand{\TwoGluonVertex}{ \makebox{\raisebox{-1.6cm}{\includegraphics{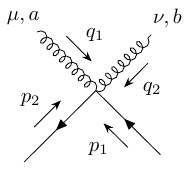}~}} }
\newcommand{\ThreeGluonVertex}{ \makebox{\raisebox{-1.6cm}{\includegraphics{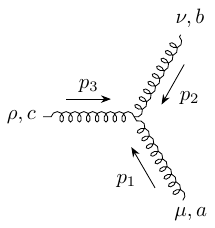}~}} }
\newcommand{\FourGluonVertex}{ \makebox{\raisebox{-1.6cm}{\includegraphics{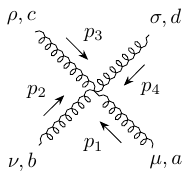}~}} }

{\catcode`\|=\active\gdef\Braket#1{\left<\mathcode`\|"8000\let|\bravert 
{#1}\right>}}

\def\bravert{\egroup\,\vrule\,\bgroup}

\usepackage{color}
\usepackage{amssymb,amsmath,amsfonts}
\usepackage{epstopdf} 
\usepackage{braket}

\usepackage{tikz-feynman}
\tikzfeynmanset{compat=1.0.0}

\usepackage{autobreak}

\begin{document}
%
%
\title{\vspace*{0.5in} 
Dijet spectrum in nonlocal and asymptotically nonlocal theories 
\vskip 0.1in}
\author{Mikkie R. Anderson}\email[]{mrmusser@wm.edu}
\author{Christopher D. Carone}\email[]{cdcaro@wm.edu}
\affiliation{High Energy Theory Group, Department of Physics,
William \& Mary, Williamsburg, VA 23187-8795, USA} 
\date{June 17, 2024}
\begin{abstract}
Asymptotically nonlocal field theories approximate ghost-free nonlocal theories at low energies, yet are theories of finite order in the number of derivatives.  These theories have an emergent nonlocal scale that regulates loop diagrams and can provide a solution to the hierarchy problem.  Asymptotic nonlocality has been studied previously in scalar theories, Abelian and non-Abelian gauge theories with complex scalars, and linearized gravity.  Here we extend that work by considering an asymptotically nonlocal generalization of QCD, which can be used for realistic phenomenological investigations.  In particular, we derive Feynman rules relevant for the study of the production of dijets at hadron colliders and compute the parton-level cross sections at leading order.   We use these to determine a bound on the scale of new physics from Large Hadron Collider data, both for a typical choice of model parameters, and in the nonlocal limit.
\end{abstract}

\maketitle
\newpage
\section{Introduction and Framework}\label{sec:intro}

The Lee-Wick Standard Model (LWSM) is a theory with higher-derivative quadratic terms, leading to propagators that fall off more quickly with momentum than those of the Standard Model~\cite{Grinstein:2007mp}.  As a consequence, the quadratic divergence of the Higgs boson squared mass is eliminated and the hierarchy problem is resolved.   Each propagator in the LWSM has an additional pole representing a new, heavy particle that is a ``partner" to the given Standard Model particle.   The residues of the new poles are opposite in sign to those of ordinary particles; in an auxiliary field description, this sign difference leads to diagrammatic cancellations that reproduce the expected ultraviolet behavior of the higher-derivative theory.  Wrong-sign residues imply that Lee-Wick particles are ghosts.  Nevertheless, it has been argued that if Lee-Wick particles are excluded from the spectrum of asymptotic scattering states, and if loop diagrams are evaluated using appropriate pole 
prescriptions~\cite{LeeWick:1969,Cutkosky:1969fq,Anselmi}, Lee-Wick theories are unitary and viable as extensions of the Standard Model.   

The LWSM, like the minimal supersymmetric extension of the Standard Model, predicts heavy particles that have not been observed.  While new particle masses can always be pushed just above current experimental bounds, doing so gradually reintroduces the unwanted fine-tuning needed to keep the Higgs boson mass close to the weak scale.  While the precise amount of fine-tuning that is tolerable may be debated, the reintroduction of fine tuning motivates consideration of higher-derivative theories that do not predict unobserved heavy particles at the TeV scale.

Nonlocal theories present such a possibility (see, for example, Refs.~\cite{Efimov:1967,Krasnikov:1987,Kuzmin:1989,Tomboulis:1997gg,Modesto:2011kw,Biswas:2011ar,Buoninfante:2018mre}).  In these theories, the mass and kinetic terms in the Lagrangian are typically modified by a nonlocal form factor, an infinite-derivative operator that is an entire function of  $\Box/\Lambda_{\rm nl}^2$, where $\Box \equiv \partial_\mu \partial^\mu$ and $\Lambda_{\rm nl}$ is the nonlocal scale.  Such a choice modifies the ultraviolet behavior of propagators without introducing additional poles.   The simplest constructions have employed the exponential of the $\Box$ operator, as in this generalization of the theory of a real scalar field:
\begin{equation}\label{eq:example}
\mathcal{L}_\infty = -\frac{1}{2} \phi \, (\Box + m_\phi^2)\,  e^{\ell^2 \Box } \, \phi - V(\phi) \,\,\, .
\end{equation}
Here $\ell \equiv 1/\Lambda_{\rm nl}$.  The $\phi$ propagator involves a factor of $e^{\ell^2 p^2}$ which becomes $e^{-\ell^2 p_E^2}$ in loop amplitudes after Wick rotation, where $p_E$ is the Euclidean momentum.   This leads to improved convergence, with $\Lambda_{\rm nl}$ serving as a regulator scale.    

Asymptotically nonlocal theories represent another possibility, one that interpolates between Lee-Wick theories and ghost-free nonlocal theories~\cite{Boos:2021chb,Boos:2021jih,Boos:2021lsj,Boos:2022biz,Carone:2023cnp}.  These theories allow the decoupling of the Lee-Wick particles without reintroducing the fine-tuning problem due to the emergence of a derived regulator scale ({\it i.e.}, one that does not appear as a fundamental parameter in the Lagrangian) that is hierarchically smaller than the lightest Lee-Wick resonance mass.  Asymptotically nonlocal theories have been explored in the recent literature in the context of scalar theories~\cite{Boos:2021chb}, Abelian gauge theories~\cite{Boos:2021jih}, non-Abelian gauge theories~\cite{Boos:2021lsj} and linearized gravity~\cite{Boos:2022biz}.     To review the basic construction, we note that Eq.~(\ref{eq:example}) is recovered from
\begin{equation} \label{eq:expapprox}
\mathcal{L} = -\frac{1}{2}\phi \, (\Box+m_\phi^2) \left( 1+\frac{\ell^2\Box}{N-1} \right)^{N-1}\!\phi  - V(\phi)~,
\end{equation}
in the limit that $N$ is taken to infinity.   At finite $N$, this theory is not quite what we want, since the $\phi$ propagator has an $(N-1)^\textrm{th}$ order pole, which does not have a simple particle interpretation.   However, we can obtain the same limiting form by working instead with 
\begin{equation}\label{eq:lagnondeg}
\mathcal{L}_N = -\frac{1}{2}\phi \, (\Box+m_\phi^2) \left[ \prod_{j=1}^{N-1}  \left( 1+\frac{\ell_j^2\Box}{N-1} \right)\right]\phi - V(\phi)~,
\end{equation}
where the $\ell_j$ are nondegenerate but approach a common value, $\ell$, as $N$ becomes large.  In this case, the propagator is given by
\begin{equation}
\label{eq:DF}
D_F(p^2) = \frac{i}{p^2-m_\phi^2} \prod_{j=1}^{N-1} \left( 1-\frac{\ell_j^2 p^2}{N-1} \right)^{-1}~,
\end{equation}
which has $N$ first-order poles, representing a spectrum of particles with masses $m_\phi$ and  $m_j \equiv \sqrt{N-1}/\ell_j$, for $j=1 \ldots N-1$.  In the past literature~\cite{Boos:2021chb,Boos:2021jih,Boos:2021lsj,Boos:2022biz,Carone:2023cnp}, a convenient parameterization was chosen for how the $m_j$ are decoupled as $N$ becomes large, while the regulator scale $\ell$ is held fixed, namely
\begin{equation}\label{eq:masses}
m_j^2 = \frac{N}{\ell^2} \frac{1}{1-\frac{j}{2 N^P}} ~, \hspace{1cm}~j=1~.~.~.~N-1 \,\,\,\,\, , \,\,\,\,\, P>1 \,\,\, .
\end{equation}
The results discussed in Refs.~\cite{Boos:2021chb,Boos:2021jih,Boos:2021lsj,Boos:2022biz,Carone:2023cnp} did not depend strongly on how the nonlocal limiting theory was approached.  For any finite $N$, the propagator, Eq.~(\ref{eq:DF}), may be expressed via a partial fraction decomposition as a sum over simple poles with residues of alternating signs (a behavior that is expected in higher-derivative theories~\cite{Pais:1950za}).  The poles with wrong-sign residues are Lee-Wick particles.  Lee-Wick theories involving higher-derivative terms that are of higher-order than those found in the LWSM have been considered before~\cite{Carone:2008iw}, including the identification of equivalent auxiliary field formulations (that is, with Lagrangians expressed in
terms of additional fields but without higher-derivative terms).   Auxiliary field formulations were also considered in the context of asymptotically nonlocal theories in Refs.~\cite{Boos:2021chb,Boos:2021jih,Boos:2021lsj,Boos:2022biz,Carone:2023cnp};  here, we work exclusively in the higher-derivative formulation of these theories.     

The propagator in Eq.~(\ref{eq:DF}) can be expressed in terms of the masses $m_j$,
\begin{equation}
D_F(p^2)=\frac{i}{(p^2-m_\phi^2)\prod_{j=1}^{N-1}(1-p^2/m_j^2)}~.
\label{eq:propnice}
\end{equation}
For Euclidean momentum, the product in the denominator of Eq.~(\ref{eq:propnice}) approaches a growing exponential in the large $N$ limit of Eq.~(\ref{eq:masses}).   This regulates loop diagrams at the scale $\Lambda_{\rm nl}$, where $\Lambda^2_{\rm nl}$ is roughly a factor of $N$ smaller than the square of the lightest Lee-Wick resonance mass $m_1^2$.

Asymptotically nonlocal theories represent a class of higher-derivative theories that are different from the simplest Lee-Wick theories and ghost-free nonlocal theories, which makes study of their properties and phenomenology well motivated.    These theories may provide a different approach to considering unitarity in nonlocal theories~\cite{NLunitarity}, namely by applying approaches that are known to work in Lee-Wick theories of finite order~\cite{LeeWick:1969,Cutkosky:1969fq,Anselmi} and then taking the limit as $N$ becomes large.   Of greater relevance to the present work is that asymptotically nonlocal theories can be considered the ultraviolet completions of theories that appear nonlocal at low energies.  Tree-level scattering processes at the Large Hadron Collider (LHC) exist in Minkowski space, where the exponential factor in Eq.~(\ref{eq:example}) may produce unbounded growth in cross sections with center-of-mass energy.  In asymptotically nonlocal theories, however, such growth is truncated due to the change in the theory at the scale of the first Lee-Wick resonance, $m_1$~\cite{Carone:2023cnp}.  In other words, if one were to integrate out all the heavy particles in an effective field theory approach, then the effective theory below the cutoff $m_1$ would look (approximately) like a ghost-free nonlocal theory;  the asymptotically nonlocal theory provides an ultraviolet completion.

From a phenomenological perspective,  it is natural to seek a bound on the nonlocal scale $\Lambda_{\rm nl}$~\cite{Biswas:2014yia}.    While asymptotically nonlocal theories delay the appearance of new particles, the momentum dependence of scattering amplitudes is nonetheless affected by the same physics that accounts for the regulation of loop diagrams which, based on naturalness arguments, one would expect to be associated with the TeV scale.
Since the LHC is currently the highest-energy collider available to probe new physics,  it is natural to investigate how the relevant physics might be probed there, in one of the most common processes:  the production of dijets.   Hence, we will focus on computing the parton-level cross sections in an asymptotically nonlocal generalization of QCD that determine the proton-proton cross section for dijet production, in particular,  the differential cross section with respect to the dijet invariant mass.  The dijet invariant mass spectrum has been used in other contexts to bound new physics, for example, to determine a lower bound on the mass of colorons  in Ref.~\cite{Bertram:1998wf}.   The Feynman rules for asymptotically nonlocal QCD have not appeared in the literature (only scalar QCD was considered in Ref.~\cite{Boos:2021lsj}), so we first determine the rules relevant to two-into-two scattering in the next section.  We then give our expressions for the parton-level cross sections $\hat{\sigma}$, which are significantly more complicated than what one obtains in QCD, and explain how gauge-fixing and the identification of asymptotic states works in our higher-derivative construction.   The expressions for the  various  $\hat{\sigma}$ also have not appeared before in the literature and can be incorporated in detailed collider physics studies.  While an exhaustive collider physics study is not the focus of the present work,  we nevertheless use our theoretical results and data from the LHC to obtain a bound on the nonlocal scale from the dijet invariant mass spectrum.  In the final section, we summarize our conclusions.

\section{Asymptotically nonlocal QCD}
An asymptotically nonlocal SU(N) gauge theory with complex scalar matter was presented in Ref.~\cite{Boos:2021lsj}, where loop corrections to the scalar two-point function were studied given their relevance to the hierarchy problem.  Here we are interested in a realistic SU(3) gauge theory with spin-$1/2$ fermions, namely QCD with color-triplet quarks, for phenomenological applications.   Following the notation of Ref.~\cite{Boos:2021lsj}, we define a covariant box operator $\underline{\Box}\equiv D_\mu D^\mu$, with SU(3) covariant derivative $D_\mu = \partial_\mu -i \, g \, T^a A_\mu^a$
and
\begin{equation} \label{eq:fdef}
f(\underline{\Box})\equiv \prod_{j=1}^{N-1}\left( 1+a_j^2\underline{\Box} \right)~,
\end{equation}
where we define  $a_j^2 \equiv \ell_j^2/(N-1)$.  Eq.~(\ref{eq:fdef}) is a gauge-covariant version of the higher-derivative product that appears in Eq.~(\ref{eq:lagnondeg}).   We then define
the asymptotically nonlocal extension of QCD by inserting $f(\underline{\Box})$ in the kinetic and mass terms, in analogy to Eq.~(\ref{eq:lagnondeg}),
\begin{equation}\label{eq:nlqcdlag}
\mathcal{L}= -\frac{1}{2}\mbox{Tr }F_{\mu\nu} f(\underline{\Box}) \, F^{\mu\nu}  + \frac{1}{2}\overline{q}\{ (i\slashed{D}-m_q),f(\underline{\Box}) \} q+ {\cal L}_{\rm g.f.}~,
\end{equation}
where ${\cal L}_{\rm g.f.}$ represents gauge-fixing terms.  Here, $F^{\mu\nu} \equiv F^{\mu\nu a} T^a$,  and the flavor indices on the quark field have been suppressed. The braces in the first term represent an anti-commutator, defined by $\{X,Y\} \equiv X \, Y+Y \, X$, which is included to preserve the hermiticity of the Lagrangian. In the local limit, $f(\underline{\Box})\rightarrow 1$, one obtains the usual QCD Lagrangian.  We assume a familiar form for the gauge-fixing term,
\begin{equation} \label{eq:gft}
 {\cal L}_{\rm g.f.} = -\frac{1}{2\xi}(\partial^\mu A_\mu^a)^2  \,\,\,.
\end{equation}
A nonlocal modification to the gauge-fixing term is unnecessary, as nothing physical depends on this choice; the form in Eq.~(\ref{eq:gft}) is convenient for implementing the usual Fadeev-Popov gauge fixing ansatz.

\subsection{Feynman rules}\label{sec:feynrules}
The quark and gluon propagators follow from the purely quadratic terms in Eqs.~(\ref{eq:nlqcdlag}) and~(\ref{eq:gft}). For the quark fields we find 
\begin{equation}
D(p)=\frac{i \, (\slashed{p}+m_q)}{(p^2-m_q^2)f(-p^2)}~,
\end{equation}
while for the gluons
\begin{equation} \label{eq:glueprop}
{D}^{ab}_{\mu\nu}(p) = -\frac{i}{p^2 f(-p^2)}\left[ \eta_{\mu\nu}-\frac{p_\mu p_\nu}{p^2}\left(1-\xi f(-p^2)\right) \right]\delta^{ab}~,
\end{equation} 
where $a$ and $b$ are color indices.  In the calculations that we present in Sec.~\ref{sec:partoncrosssec}, we will work in the nonlocal equivalent of Landau gauge, 
where $\xi=0$, as this simplifies intermediate algebraic steps.  We note that the factor of $f(-p^2)$ in the denominator of Eq.~(\ref{eq:glueprop}) becomes a 
growing exponential as a function of Euclidean momentum in the nonlocal limit, which accounts for the elimination of quadratic divergences in the theory of complex scalars 
discussed in Ref.~\cite{Boos:2021lsj}.

To evaluate the two-into-two scattering processes of interest to us, we need the interaction vertices involving at least one gluon and no more than 4 lines of any type.  It is straightforward, though somewhat tedious, to extract the interactions involving a specified number of gluon fields from the Lagrangian that involves the product of an arbitrary number of covariant box operators defined in Eq.~(\ref{eq:fdef}).   For vertices involving a quark line, one can have either one or two gluon lines.  We find the Feynman rules
\begin{equation}
\OneGluonVertex
= i \, g \, T^a \, V_{1g}^\mu(p_1,p_2) \, , 
\end{equation}
where
\begin{equation}
V_{1g}^\mu (p_1,p_2) \equiv \frac{1}{2} \left[ f_1(p_1^2)+f_1(p_2^2) \right]\gamma^\mu  -(p_1-p_2)^\mu \left( \frac{\slashed{p}_1-\slashed{p}_2}{2}-m_q \right) f_2(p_1^2,p_2^2)
\,\,\, ,
\end{equation}
and
\begin{equation}
\TwoGluonVertex = -i \, g^2 \, T^aT^b \, V_{2g}^{\mu\nu}(p_1,p_2,q_1,q_2)  + \left[\{q_1,\mu,a\}\leftrightarrow \{q_2,\nu,b\}\right] ,  \,\,\, 
\end{equation}
where
\begin{eqnarray}
 &&V_{2g}^{\mu\nu} (p_1,p_2,q_1,q_2) \equiv  \eta^{\mu\nu} \left( \frac{\slashed{p}_1-\slashed{p}_2}{2}-m_q \right) f_2(p_2^2,p_1^2)  
 + (q_1+2 \, p_2)^\mu(q_2+2 \, p_1)^\nu   \nonumber \\   &&\times \left( \frac{\slashed{p}_1-\slashed{p}_2}{2}-m_q \right) f_3(p_2^2,(q_2+p_1)^2,p_1^2) + \frac{1}{2}\gamma^\mu (q_2+2 \,p_1)^\nu f_2((q_2+p_1)^2,p_1^2) \nonumber \\
&&-\frac{1}{2}(q_1+2\, p_2)^\mu \gamma^\nu f_2(p_2^2,(q_2+p_1)^2)  \,\,\, .
 \end{eqnarray} 
 
The three- and four-gluon self-interactions are the same as those found in Ref.~\cite{Boos:2021lsj}.  We provide these Feynman rules here for completeness:
\begin{equation} \label{eq:threegfr}
\ThreeGluonVertex
= -g\, f^{abc} \, V_{3g}^{\mu\nu\rho} (p_1,p_2,p_3) + \mbox{ all permutations} , 
\end{equation}
where
\begin{equation} \label{eq:threegfr2}
V_{3g}^{\mu\nu\rho} (p_1,p_2,p_3) \equiv \eta^{\mu\rho}p_1^\nu f_1(p_1^2)+\frac{1}{2}(p_1-p_3)^\nu(p_1\cdot p_3 \, \eta^{\mu\rho}-p_1^\rho p_3^\mu)\,f_2(p_1^2,p_3^2) \,\,\, .
\end{equation}
Here ``all permutations" refers to the $3!$ ways we may permute the elements of the set $\{ (p_1,\mu, a), \, (p_2,\nu,b), \, (p_3,\rho,c)\}$, which
label the lines of the vertex.  Finally,
\begin{equation}\label{eq:fourgfr}
\FourGluonVertex
= - i g^2 f^{abe}f^{cde} V_{4g}^{\mu\nu\rho\sigma} (p_1,p_2,p_3,p_4) + \mbox{ all permutations} , 
\end{equation}
where
\begin{eqnarray} \label{eq:fourgfr2}
V_{4g}^{\mu\nu\rho\sigma} (p_1,p_2,p_3,p_4) = \frac{1}{4}\eta^{\mu\rho}\eta^{\nu\sigma} f_1\!\left((p_3+p_4)^2\right) &-& \eta^{\nu\sigma} p_4^\mu (p_3+2 \, p_4)^\rho f_2\!\left((p_1+p_2)^2,p_4^2\right)  \nonumber \\ 
- \frac{1}{2} \eta^{\nu\rho}(p_1\cdot p_4 \, \eta^{\mu\sigma}-p_1^\sigma p_4^\mu)\,f_2(p_1^2,p_4^2)& -&\frac{1}{2} (2 \, p_1+p_2)^\nu (p_3+2\,p_4)^\rho   \nonumber \\ \times(p_1\cdot p_4 \, \eta^{\mu\sigma}-p_1^\sigma p_4^\mu) &&\hspace{-1em} f_3\!\left(p_1^2,(p_1+p_2)^2,p_4^2\right)  \,\,\, .
\end{eqnarray}
In these Feynman rules, we define the functions $f_1$, $f_2$ and $f_3$ as follows: 
\begin{equation}
\begin{split}
f_1(p^2)&\equiv \prod_{j=1}^{N-1} (1-a_j^2p^2) \,\, ,\\
f_2(p_1^2,p_2^2)&\equiv \sum_{k=1}^{N-1} a_k^2\left[ \prod_{j=1}^{k-1}(1-a_j^2p_1^2) \right]\left[ \prod_{j=k+1}^{N-1} (1-a_j^2p_2^2) \right] 
\,\, ,\\
f_3(p_1^2,p_2^2,p_3^2)&\equiv \sum_{n=1}^{N-1}\sum_{k=n+1}^{N-1}a_n^2a_k^2\left[ \prod_{j=1}^{n-1}(1-a_j^2p_1^2) \right]\left[ \prod_{j=n+1}^{k-1}(1-a_j^2p_2^2) \right] \left[ \prod_{j=k+1}^{N-1}(1-a_j^2p_3^2) \right]~.
\end{split}
\end{equation}
As one might surmise, the functions $f_2$ and $f_3$ arise by extracting the one- and two-gluon parts of the product in Eq.~(\ref{eq:fdef}), respectively.   As noted in Ref.~\cite{Boos:2021lsj}, these functions are totally symmetric under interchange of their arguments and approach the following exponential forms in the large $N$ limit:
\begin{equation} \label{eq:expforms}
\begin{split}
\lim_{N\rightarrow \infty} f_1(p^2) &= e^{-\ell^2p^2} \,\, ,\\
\lim_{N\rightarrow\infty} f_2(p_1^2,p_2^2) &= \frac{e^{-\ell^2p_1^2}-e^{-\ell^2p_2^2}}{p_2^2-p_1^2} \,\, ,\\
\lim_{N\rightarrow \infty} f_3(p_1^2,p_2^2,p_3^2) &= \frac{e^{-\ell^2p_1^2}}{(p_2^2-p_1^2)(p_3^2-p_1^2)}+\frac{e^{-\ell^2p_2^2}}{(p_1^2-p_2^2)(p_3^2-p_2^2)}+\frac{e^{-\ell^2p_3^2}}{(p_1^2-p_3^2)(p_2^2-p_3^2)}~.
\end{split}
\end{equation}
In the limit that $\Lambda_{nl}\rightarrow\infty$, the $a_k \rightarrow 0$, so that $f_1(p^2)\rightarrow 1$, $f_2(p_1^2,p_2^2) \rightarrow 0$
and $f_3(p_1^2,p_2^2,p_3^2) \rightarrow 0$, independent of the arguments of these functions and the value of $N$. One thereby recovers the QCD Lagrangian
in this limit.

\subsection{Two-into-two parton-level cross sections}\label{sec:partoncrosssec}
Following the notation of Ref.~\cite{Eichten:1984eu}, the cross-section for a two-jet final state can be expressed as 
\begin{align}
\frac{d\sigma}{d y_1 dy_2 dp_\perp} =  \frac{2 \pi}{s} p_\perp & \sum_{ij} \left[
f_i^{(a)}(x_a,Q^2) f_j^{(b)}(x_b,Q^2) \, \hat{\sigma}_{ij}(\hat{s},\hat{t},\hat{u})  \right. \nonumber \\
& \left. + f_j^{(a)}(x_a,Q^2) f_i^{(b)}(x_b,Q^2) \, \hat{\sigma}_{ij}(\hat{s},\hat{u},\hat{t}) \right]
/(1+\delta_{ij}) \,\,\, ,
\label{eq:defsighat}
\end{align} 
where $y_1$ and $y_2$ are the jet rapidities, $p_\perp$ is the jet transverse momentum, the $f_i$ are parton distribution functions, and $s$, $t$, and $u$ are the Mandelstam variables with a hat indicating those of the parton-level process.  We comment further on
the kinematical variables that are relevant to our later analysis and on the arguments of the parton distribution functions in Sec.~\ref{sec:bound}.    Here, we simply note that Eq.~(\ref{eq:defsighat}) defines the parton-level cross sections 
$\hat{\sigma}_{ij}$, which have been known for some time in QCD but are modified in the asymptotically nonlocal theories we consider here. In this section and in an Appendix, we summarize the results we obtain for the $\hat{\sigma}$, which were computed using the Feynman rules of Sec.~\ref{sec:feynrules} via the FeynCalc package~\cite{FeynCalc} in Mathematica.  

Before proceeding to these results, we make a few technical comments.  First, we note that a field in the higher-derivative theory is associated with a number of distinct particle states, while we are interested in diagrams where the external lines correspond to the lightest of these states.   As described in Refs~\cite{Boos:2021chb,Boos:2021jih,Boos:2021lsj,Boos:2022biz,Carone:2023cnp}, a higher-derivative field can be decomposed into a sum of quantum fields in an auxiliary field description where each exclusively creates or annihilates one type of particle.  The coefficient of the component field that annihilates or creates the lightest state is determined by the wave function renormalization factor that one finds at the corresponding pole in the higher-derivative theory.   For massless partons, the form of our Lagrangian assures that this factor is unity [since $f(0)=f_1(0)=1$], so that the field in the higher-derivative theory creates or annihilates the lightest particle component without any numerical correction factor compared to a canonically normalized quantum field in a theory that has conventional mass and kinetic terms.  Secondly, we mentioned earlier that we work in the higher-derivative generalization of Landau gauge, which implies that we must include ghosts if we sum over all possible polarization states of the external gluon lines.   Alternately, we may omit the ghosts if we also omit the unphysical polarization states that the ghosts would cancel in the polarization sums.  This can be accomplished using standard techniques involving an auxiliary vector (see, for example, Sec.~3 of Ref.~\cite{schwinn}).  This is the approach we follow and we have verified as a consistency check that our cross sections correctly reproduce all the expected QCD results in the limit that the scale of new physics is taken to be infinitely large. 

For the case of quark-antiquark annihilation through $s$-channel gluon exchange, the cross section is given by
\begin{equation}
\hat{\sigma}_{q_i\overline{q}_i\rightarrow q_j\overline{q}_j} = \frac{4 \, \alpha_s^2}{ 9\, \hat{s}}  \frac{\hat{t}^2+\hat{u}^2}{\hat{s}^2 f_1(\hat{s})^2}~,
\,\,\,\,\, i \neq j ,
\end{equation}
where $i$ and $j$ are quark flavor indices.   Here, and henceforth, we assume all partons are massless, and the final state jets include five light flavors, with the top quark excluded.  For $t$-channel scattering of different flavors of quark
or antiquark, the cross section is
\begin{equation}
\hat{\sigma}_{q_i q_j \rightarrow q_i q_j} = \frac{4 \, \alpha_s^2}{ 9\, \hat{s}}    \frac{\hat{s}^2+\hat{u}^2}{\hat{t}^2f_1(\hat{t})^2}~, \,\,\,\,\, i \neq j .
\end{equation} 
For the special case of quark-antiquark scattering into quark-antiquark of the same flavor, there are both $s$- and $t$-channel contributions
\begin{equation}
\hat{\sigma}_{q_i\overline{q}_i\rightarrow q_i\overline{q}_i} = \frac{4 \, \alpha_s^2}{ 9\, \hat{s}} 
\left( \frac{\hat{t}^2+\hat{u}^2}{\hat{s}^2 f_1(\hat{s})^2} + \frac{\hat{s}^2+\hat{u}^2}{\hat{t}^2 f_1(\hat{t})^2} - \frac{2\hat{u}^2}{3\hat{s}\hat{t}f_1(\hat{s})f_1(\hat{t})} \right)~,
\end{equation}
and for the similar case of quark-quark scattering of a single flavor, there are $t$- and $u$-channel diagrams, leading to
\begin{equation}
\hat{\sigma}_{q_iq_i\rightarrow q_iq_i} =  \frac{4 \, \alpha_s^2}{ 9\, \hat{s}} 
\left( \frac{\hat{s}^2+\hat{u}^2}{\hat{t}^2 f_1(\hat{t})^2} + \frac{\hat{s}^2+\hat{t}^2}{\hat{u}^2 f_1(\hat{u})^2} - \frac{2\hat{s}^2}{3\hat{t}\hat{u}f_1(\hat{t})f_1(\hat{u})} \right)~.
\end{equation}

While the modified form of the $\hat{\sigma}$ for processes exclusively involving quarks and/or antiquarks might be easy to intuit, those involving gluon external lines are much more complicated due to the modification of the Feynman rules in 
Eqs.~(\ref{eq:threegfr})-(\ref{eq:fourgfr2}).  The cross section for a quark-antiquark pair scattering into two gluons may be expressed in
the form
\begin{equation} \label{eq:qqbggform}
\begin{split}
\hat{\sigma}_{q\overline{q}\rightarrow gg} = 
\frac{ \alpha_s^2}{ 9\, \hat{s}} 
\sum_{i,j,k=0}^4 f_2(0,0)^i f_2(\hat{t},0)^j f_2(\hat{u},0)^k F_{ijk}(\hat{s},\hat{t},\hat{u})~,
\end{split}
\end{equation}
where the coefficients $F_{ijk}(\hat{s},\hat{t},\hat{u})$ are given in Appendix~\ref{sec:appendixqg}.  The function $f_2$ vanishes in the $\Lambda_{\rm nl} \rightarrow \infty$ limit, which implies that the QCD result lives entirely in the $F_{000}$ part
of Eq.~(\ref{eq:qqbggform}) in the same limit.   The parton-level cross sections $\hat{\sigma}_{gg\rightarrow q\overline{q}}$ and $\hat{\sigma}_{qg\rightarrow qg}$ can be obtained from Eq.~(\ref{eq:qqbggform}) by means of crossing symmetry.  This involves specific interchanges of Mandelstam variables, as well as adjustments in overall signs and spin/color factors, as discussed in standard textbooks~\cite{Peskin:1995ev}.  We find
\begin{equation} \label{eq:ggqqbcross}
\hat{\sigma}_{gg\rightarrow q\overline{q}} = \frac{9}{64} \,\hat{\sigma}_{q\overline{q}\rightarrow gg}(\hat{t}\leftrightarrow \hat{u})~.
\end{equation}
and
\begin{equation} \label{eq:qgqgcross}
\hat{\sigma}_{qg\rightarrow qg} =\hat{\sigma}_{\overline{q}g\rightarrow \overline{q}g}  = -\frac{3}{8} \, \hat{\sigma}_{q\overline{q}\rightarrow gg} (\hat{s}\leftrightarrow \hat{t})~,
\end{equation}

Finally, the cross section for gluon-gluon scattering to two gluons may be written in the form
\begin{equation}
\hat{\sigma}_{gg\rightarrow gg} =
\frac{\alpha_s^2}{\hat{s}} 
 \sum_{i,j,k,\ell,m=0}^4 f_2(0,0)^i f_2(\hat{t},0)^j f_2(\hat{u},0)^k f_3(0,\hat{t},0)^\ell f_3(0,\hat{u},0)^m F_{ijk\ell m}(\hat{s},\hat{t},\hat{u})~,
\end{equation}
where the coefficients $F_{ijk\ell m}(\hat{s},\hat{t},\hat{u})$ are provided in Appendix~\ref{sec:appendixgg}.  Again, the QCD limit lives entirely in the term involving $F_{00000}(\hat{s},\hat{t},\hat{u})$.\footnote{A Mathematica file with all the $\hat{\sigma}$ used in our analysis is available upon request.} 

\section{A bound from the dijet invariant mass spectrum}\label{sec:bound}
With the parton-level cross sections $\hat{\sigma}$ defined in the previous section, we may compute the cross section for 
$p\,p \rightarrow \mbox{jet jet}$ with the goal of determining a bound on the nonlocal scale 
$\Lambda_{\rm nl}$ using LHC data.  We focus on the dijet invariant mass spectrum which is related to the $\hat{\sigma}$ via
\begin{align}
\frac{d\sigma}{d{\cal M}} = & \frac{\pi \, {\cal M}}{2 \, s} \int_{-Y}^Y  dy_1\, \int_{y_{\rm min}}^{y_{\rm max}} dy_2 \mbox{ sech}^2y_* \, \sum_{ij} \left[
f_i^{(a)}(x_a,Q^2) f_j^{(b)}(x_b,Q^2) \, \hat{\sigma}_{ij}(\hat{s},\hat{t},\hat{u})  \right. \nonumber \\
& \left. + f_j^{(a)}(x_a,Q^2) f_i^{(b)}(x_b,Q^2) \, \hat{\sigma}_{ij}(\hat{s},\hat{u},\hat{t}) \right]
/(1+\delta_{ij}) \,\,\, .
\label{eq:partprot}
\end{align} 
Here ${\cal M}$ is the dijet invariant mass, the $y_i$ are the jet rapidities in the proton-proton center of mass frame, with  the boost-invariant quantity $y_* \equiv (y_1-y_2)/2$.  Since we treat the partons as massless, there is no distinction between rapidty and pseudorapidity, so we use these terms interchangeably. The parton distribution function for the $i^{\rm th}$ parton within hadron $a$,  $f_i^{(a)}(x_a,Q^2)$, is a function of the parton momentum fraction $x_a$ and the renormalization scale $Q$.  The Mandelstam variables $\hat{s}$, $\hat{t}$ and $\hat{u}$, and the momentum fractions $x_a$ and $x_b$,  are related to ${\cal M}$ and the 
integration variables by
\begin{equation} \label{eq:shat}
\hat{s} = \, {\cal M}^2  \,\,\, ,
\end{equation}
\begin{equation} \label{eq:that}
\hat{t} = -\frac{1}{2} \, {\cal M}^2 \, ( 1 - \tanh y_*) \,\,\, ,
\end{equation}
\begin{equation} \label{eq:uhat}
\hat{u} = -\frac{1}{2} \, {\cal M}^2 \, ( 1 + \tanh y_*) \,\,\, ,
\end{equation}
\begin{equation} \label{eq:xa}
x_a = \frac{{\cal M}}{\sqrt{s}} \, e^{\, y_{\rm boost}}   \,\,\, ,
\end{equation}
\begin{equation} \label{eq:xb}
x_b = \frac{{\cal M}}{\sqrt{s}}\,  e^{-y_{\rm boost}} \,\,\, ,
\end{equation}
where $y_{\rm boost} \equiv (y_1+y_2)/2$ and $\sqrt{s}$ is the proton-proton center-of-mass energy.   The proton-proton cross section in Eq.~(\ref{eq:partprot}) assumes a cut $Y>0$ is placed on the jet rapidity, such that $|y_i|<Y$; this leads to the integration region shown with
\begin{align}
y_{\rm min} & = \mbox{max }(-Y, \ln \tau - y_1) \,\,\, ,\label{eq:minmax} \\
y_{\rm max} & = \mbox{max }(Y,- \ln \tau - y_1) \,\,\ ,\label{eq:minmax2}
\end{align}
where $\tau = {\cal M}^2 / s$.  Eqs.~(\ref{eq:minmax}) and (\ref{eq:minmax2}) follow from the allowed range of the momenta fractions $x_a$ and $x_b$ which must fall between $0$ and $1$.  
Note that Eqs.~(\ref{eq:partprot})-(\ref{eq:minmax2}) are well established and can be found in the literature on hadron collider physics, for example, in 
Ref.~\cite{Eichten:1984eu}.

We wish to compare the predictions of our scenario with data on the dijet invariant mass spectrum from the LHC.   The dijet spectrum has been considered in searches for new, heavy resonances (see, for example, Refs~\cite{CMS:2019gwf,CMS:2018mgb,ATLAS:2018qto}) providing us with experimental results that we can utilize to determine a bound in the present scenario.  For definiteness, we use the results from the CMS experiment that are displayed in Fig.~5 of Ref.~\cite{CMS:2019gwf}.   To match this data, we assume a rapidity cut of $Y=2.5$;  Ref.~\cite{CMS:2019gwf} places an additional cut on the difference between the pseudorapidities, translating to $|y_1-y_2|<1.1$, which we impose by including an appropriate Heaviside theta-function in the integrand of Eq.~(\ref{eq:partprot}) that vanishes when this constraint is not satisfied.   To compare to this data set, we set the proton-proton center of mass energy $\sqrt{s}=13$~TeV, and evaluate the dijet spectrum over the range $1.5$~TeV$ \leq  {\cal M} \leq 8.5$~TeV, with the renormalization scale $Q$ set equal to the dijet invariant mass ${\cal M}$.   Eq.~(\ref{eq:partprot}) is evaluated numerically on Mathematica using the ManeParse package~\cite{Clark:2016jgm} which provides convenient access to parton distribution functions (pdfs)~\cite{Kovarik:2015cma}.  We used the nCTEQ15 pdfs for free protons in this computation. We normalize our theoretical prediction for a given nonlocal scale $\Lambda_{\rm nl}$ to the result that is obtained when the nonlocal scale is taken to infinity, {\it i.e.}, setting $f_1=1$ and $f_2=f_3=0$.   We compare this to the same ratio of data to QCD prediction given in  Ref.~\cite{CMS:2019gwf}. 

\begin{figure}[t]
\includegraphics[width=.5\textwidth]{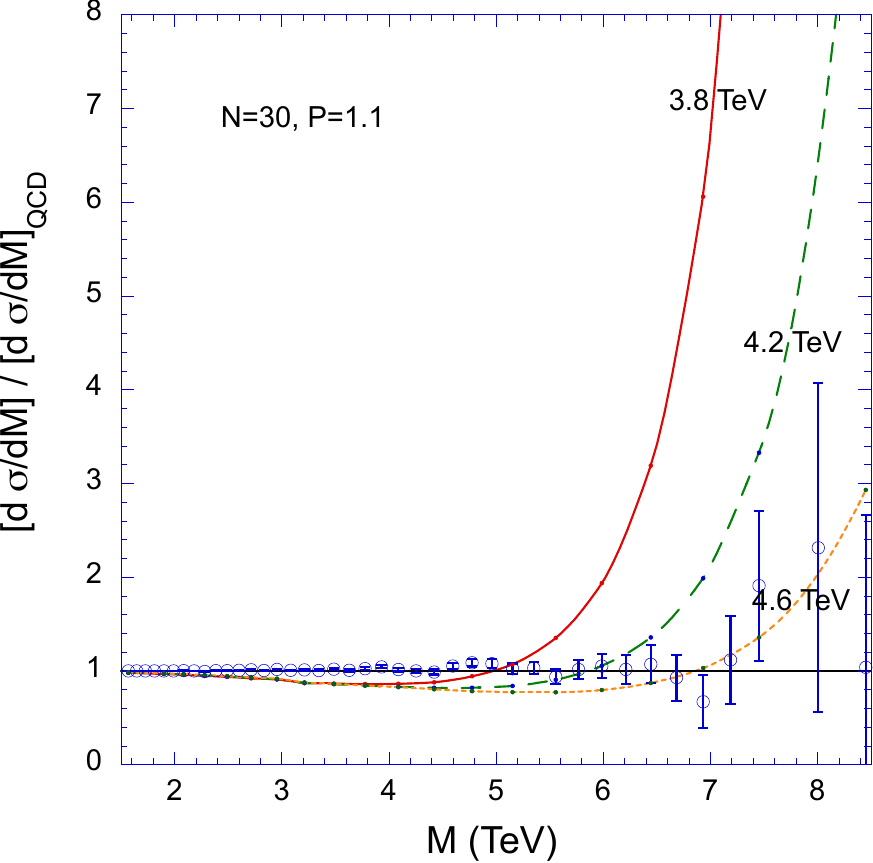}
\caption{Ratio of the predicted dijet invariant mass spectrum to the Standard Model expectation, for $N=30$, $P=1.1$ and $\Lambda_{\rm nl}=3.8$, $4.2$ and $4.6$~TeV.   The open circles represent LHC data from Ref.~\cite{CMS:2019gwf}.}  \label{fig:ratio30}
\end{figure}

\begin{figure}[t]
\includegraphics[width=.5\textwidth]{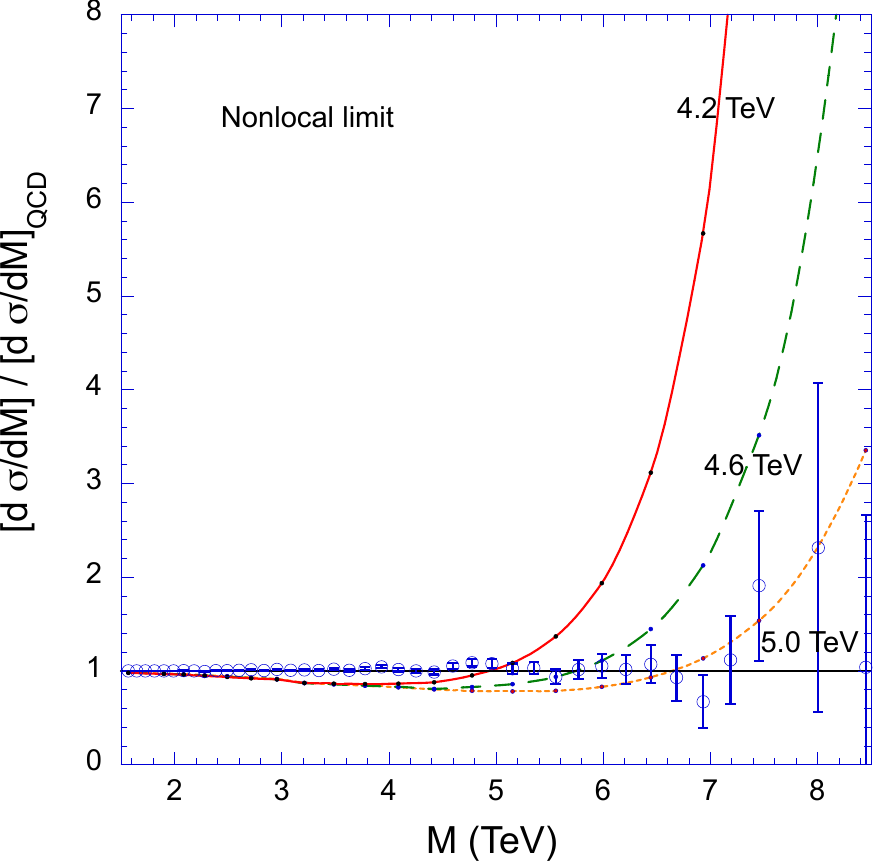}
\caption{Ratio of the predicted dijet invariant mass spectrum to the Standard Model expectation, for the nonlocal limit $N\rightarrow \infty$,
for $\Lambda_{\rm nl}=4.2$, $4.6$ and $5.0$~TeV.   The open circles represent LHC data from Ref.~\cite{CMS:2019gwf}.}  \label{fig:ratioNL}
\end{figure}

As an example of typical results, we show in Fig.~\ref{fig:ratio30} the case where there are $N=30$ poles, with $P=1.1$ in the parameterization given by Eq.~(\ref{eq:masses}), for $\Lambda_{\rm nl}=3.8$, $4.2$ and $4.6$~TeV.  The theoretical predictions shown in the figures are computed at leading order, as no computation of next-to-leading-order (NLO) effects exists for the nonlocal theory.  We assume these effects are captured by $20\%$ theoretical errors, which are comparable in size to NLO effects that have been studied in QCD (see, for example, Ref.~\cite{Frederix:2016ost}).  To determine a bound, we compute a $\chi^2$ that captures the agreement between the theoretical prediction and the data points, with total error for each data point in the $\chi^2$ function determined by adding the experimental and the assumed theoretical errors in quadrature.  We find for the case shown in Fig.~\ref{fig:ratio30} that
\begin{equation}
\Lambda_{\rm nl}^{30} > 4.2 \mbox{ TeV} \,\,\,\,\, (95\%\mbox{ C.L.}),
\label{eq:bound30}
\end{equation}
where the superscript on $\Lambda_{\rm nl}$ denotes the number of poles $N$.  We do not find that the bound differs appreciably as we vary $N$, since this parameter does not have to be very large before $f_1$, $f_2$ and $f_3$ approach their $N \rightarrow \infty$ limiting forms.   We can compute the results in the nonlocal limit using those limiting forms, given in Eq.~(\ref{eq:expforms}), which lead to Fig.~\ref{fig:ratioNL}.
In this case, the same procedure for determining a bound on the nonlocal scale gives
\begin{equation}
\Lambda_{\rm nl}^{\infty} > 4.7 \mbox{ TeV} \,\,\,\,\, (95\%\mbox{ C.L.}).
\label{eq:boundinfty}
\end{equation}
As a consistency check, we computed the same bound using the CTEQ 6.1 pdfs and found a qualitatively similar result, $\Lambda_{\rm nl}^{\infty} > 4.9 \mbox{ TeV}$
(95\% C.L.).  We note that the choices of $\Lambda_{\rm nl}$ for the curves displayed in Figs.~\ref{fig:ratio30} and \ref{fig:ratioNL} were selected to be near the bounds in Eqs.~(\ref{eq:bound30}) and (\ref{eq:boundinfty}), respectively.

We view the results of this section at finite $N$ as illustrative and similar in spirit to the analysis of the bounds on coloron models presented in Ref.~\cite{Bertram:1998wf}.  Our results assume a particular parameterization of resonance masses, namely Eq.~(\ref{eq:masses}), but the value of the theoretical results presented in our earlier sections is that they can be applied to any desried parameterization leading to different forms for the functions $f_1$, $f_2$ and $f_3$;  all should approach the same $N \rightarrow \infty$ limit.    These general results can also be used in  more detailed collider physics investigations, including realistic modeling of jets (for example, jet cone algorithms), detector acceptances and efficiencies, and studies of jet angular distributions.  Those topics go beyond the scope of the present work, and may be better motivated after a calculation of NLO effects in the nonlocal theory are at hand.

\section{Conclusions}
In this paper, we have built upon earlier work on asymptotically nonlocal field theories.  These theories appear nonlocal at low energies but have sensible ultraviolet completions in terms of Lee-Wick theories that are finite order in derivatives.  We focused on the strongly interacting sector~\cite{Boos:2021lsj}, whose modification affects the physics of jets at the highest energy hadron colliders; our goal was to obtain preliminary bounds on the scale of new physics, $\Lambda_{\rm nl}$, and provide the necessary tools for future collider analyses.    We began by determining the relevant Feynman rules for an asymptotically nonlocal SU(3) theory of fermions, since the past literature only considered a theory with complex scalar matter~\cite{Boos:2021lsj}.   While the gluon self-interactions and the procedure for gauge-fixing to obtain the gluon propagator are the same as those given in Ref.~\cite{Boos:2021lsj},  the one- and two-gluon vertices involving fermions were not previously available in the literature.  With the complete set of Feynman rules in hand, we considered the most basic jet process, dijet production from two-into-two parton scattering.   We found that the relevant parton-level cross sections are in some cases considerably more complicated than those in ordinary QCD.  Nevertheless, we checked that in the limit $ \Lambda_{\rm nl} \rightarrow \infty$,  we precisely recover the QCD results we expect in the absence of new physics.   We then computed the dijet invariant mass spectrum in proton-proton collisions at $\sqrt{s}=13$~TeV, to compare the deviation from the QCD expectation at high dijet invariant mass with experimental data from the LHC.    We found that in the exactly nonlocal limit (where the number of resonances $N$ in the asymptotically nonlocal theory is taken to infinity), the scale of new physics was bounded by 
$\Lambda_{\rm nl}>4.7$~TeV at the 95\% confidence level.   For finite $N$, we obtain bounds that are similar in magnitude, but that depend in detail on the parameterization of the Lee-Wick mass spectrum.   We presented one example with $N=30$ where we found $\Lambda_{\rm nl}>4.2$~TeV (95\% C.L.).  These bounds are similar in magnitude to other collider bounds on nonlocal theories that have been discussed in the literature~\cite{Biswas:2014yia}.

Our approach to obtaining a bound at leading order on the scale of new physics from the dijet invariant mass spectrum is similar in spirit to the bound on the coloron mass in Ref.~\cite{Bertram:1998wf}. More detailed leading-order studies might include modeling of jet hadronization, detector acceptances and efficiencies, and the effect of new physics on the angular dependence of jet cross sections.  The theoretical results presented  here make such studies feasible, but they go beyond the scope of the present work.    A more accurate assessment of the bounds on the nonlocal scale would require the computation of next-to-leading-order (NLO) effects that are not known in the asymptotically nonlocal or nonlocal theories; these have been taken into account in our assumed theoretical error bars.  A full NLO calculation in the present framework would no doubt be a complicated undertaking; it may be sensible to defer such a task until some indication of a deviation from the QCD expectations is observed at high dijet invariant masses.

\begin{acknowledgments} 
We thank the NSF for support under Grant PHY-2112460.  We thank Greg Landsberg for pointing us towards the numerical CMS data that was plotted in Ref.~\cite{CMS:2019gwf}.
\end{acknowledgments}

\appendix
\section{Full expressions}

\subsection{$q \overline{q} \rightarrow gg$ scattering amplitude}\label{sec:appendixqg}
The parton-level cross section $\hat{\sigma}_{q\overline{q}\rightarrow gg}$ was written in Sec.~\ref{sec:partoncrosssec} in the form
\begin{equation}
\begin{split}
\hat{\sigma}_{q\overline{q}\rightarrow gg} =
\frac{\alpha_s^2}{ 9\, \hat{s}} 
 \sum_{i,j,k=0}^4 f_2(0,0)^i f_2(\hat{t},0)^j f_2(\hat{u},0)^k F_{ijk}(\hat{s},\hat{t},\hat{u})~.
\end{split}
\end{equation}
The cross section $\hat{\sigma}_{gg\rightarrow q\overline{q}}$ and $\hat{\sigma}_{qg\rightarrow qg} = \hat{\sigma}_{\bar{q}g\rightarrow \bar{q}g}$ were then related to this result by crossing symmetry, in Eqs.~(\ref{eq:ggqqbcross})  and~(\ref{eq:qgqgcross}), respectively. In this appendix, we present the functions $F_{ijk}(\hat{s},\hat{t},\hat{u})$.  For each $F_{ijk}$ that we display, there is another non-vanishing one, $F_{ikj}$, found by swapping the 
$\hat{t}$ and $\hat{u}$ variables:
\begin{equation} \label{eq:swapjk}
F_{ikj}(\hat{s},\hat{t},\hat{u}) = F_{ijk}(\hat{s},\hat{u},\hat{t})~.
\end{equation}
Any coefficients not listed below, or obtained from those shown by Eq.~(\ref{eq:swapjk}), are zero.  We find:

\begin{equation}
F_{200}(\hat{s},\hat{t},\hat{u}) = \frac{12 \hat{t} \hat{u}}{f_1(\hat{s})^2} \,\,\, ,
\end{equation}

\begin{equation}
F_{022}(\hat{s},\hat{t},\hat{u}) = \frac{2\hat{t}^3\hat{u}^3}{3\hat{s}^2f_1(\hat{t})f_1(\hat{u})} \,\,\, ,
\end{equation}

\begin{equation}
F_{040}(\hat{s},\hat{t},\hat{u}) = \frac{8\hat{t}^3\hat{u}^3}{3\hat{s}^2f_1(\hat{t})^2} \,\,\, ,
\end{equation}

\begin{equation}
F_{030}(\hat{s},\hat{t},\hat{u}) = -\frac{16 \hat{t}^2\hat{u}^2(f_1(\hat{t})-1)(\hat{t}-\hat{u})}{3\hat{s}^2f_1(\hat{t})^2} \,\,\, ,
\end{equation}

\begin{equation}
F_{120}(\hat{s},\hat{t},\hat{u}) = \frac{6\hat{t}^2\hat{u}^2}{\hat{s}f_1(\hat{s})f_1(\hat{t})} \,\,\, ,
\end{equation}

\begin{equation}
F_{021}(\hat{s},\hat{t},\hat{u}) = \frac{2\hat{t}^2\hat{u}^2(f_1(\hat{u})-1)(\hat{t}-\hat{u})}{3\hat{s}^2f_1(\hat{t})f_1(\hat{u})} \,\,\, ,
\end{equation}

\begin{equation}
F_{110}(\hat{s},\hat{t},\hat{u}) = -\frac{6\hat{t}\hat{u}(f_1(\hat{t})-1)(\hat{t}-\hat{u})}{\hat{s}f_1(\hat{s})f_1(\hat{t})} \,\,\, ,
\end{equation}

\begin{equation}
F_{011}(\hat{s},\hat{t},\hat{u}) = -\frac{\hat{t}\hat{u}(f_1(\hat{t})-1)(f_1(\hat{u})-1)(3\hat{t}^2-2\hat{t}\hat{u}+3\hat{u}^2)}{3\hat{s} ^2f_1(\hat{t})f_1(\hat{u})} \,\,\, ,
\end{equation}

\begin{equation}
F_{100}(\hat{s},\hat{t},\hat{u}) =  -\frac{6\hat{t}\hat{u}}{\hat{s} f_1(\hat{s})} \left[ f_1(\hat{t})+f_1(\hat{u})+\frac{1}{f_1(\hat{t})}+\frac{1}{f_1(\hat{u})}-\frac{8}{f_1(\hat{s})} +4\right] \,\,\, ,
\end{equation}

\begin{equation}
\begin{split}
F_{010}(\hat{s},\hat{t},\hat{u}) = \frac{1}{3\hat{s}^2}\left[8\hat{u}(2\hat{t}^2-\hat{t}\hat{u}+\hat{u}^2)\left( f_1(\hat{t})-\frac{1}{f_1(\hat{t})^2} \right) + \left( \frac{1}{f_1(\hat{t})}-1 \right) \right. \\ \left.\times\left[ \hat{t}(\hat{t}^2-\hat{t}\hat{u}+2\hat{u}^2)\left( \frac{1}{f_1(\hat{u})}+f_1(\hat{u}) \right) + 2(\hat{t}^3-9\hat{t}^2\hat{u}+6\hat{t}\hat{u}^2-4\hat{u}^3)\right.\right. \\ \left.\left.+\frac{36\hat{t}\hat{u}(\hat{t}-\hat{u})}{f_1(\hat{s})}\right] \right] \,\,\, ,
\end{split}
\end{equation}

\begin{equation}
\begin{split}
F_{020}(\hat{s},\hat{t},\hat{u}) &= \frac{\hat{t}\hat{u}}{6\hat{s}^2}\left[ 8\left( 3\hat{t}^2-5\hat{t}\hat{u}+4\hat{u}^2 \right) \left( 1+\frac{1}{f_1(\hat{t})^2} \right) +\frac{1}{f_1(\hat{t})}\right. \\ &\left.\times\left[  \hat{t} f_1(\hat{u})  (\hat{t}-3\hat{u})\left( 1+\frac{1}{f_1(\hat{u})^2} \right) -2(23\hat{t}^2+11\hat{t}\hat{u}+16\hat{u}^2)+\frac{72\hat{t}\hat{u}}{f_1(\hat{s})} \right] \right] \,\,\, ,
\end{split}
\end{equation}

\begin{equation}
\begin{split}
F_{000}(\hat{s},\hat{t},\hat{u}) = \frac{1}{6\hat{s}^2}\left[\frac{288 \hat{t} \hat{u} \left(\frac{1}{f_1(\hat{s})}-1\right)}{f_1(\hat{s})}-\frac{72 \hat{t} \hat{u} \left(f_1(\hat{t})+\frac{1}{f_1(\hat{t})}+f_1(\hat{u})+\frac{1}{f_1(\hat{u})}\right)}{f_1(\hat{s})}\right. \\ \left.+\frac{4 \hat{u} \left(f_1(\hat{t})^2+\frac{1}{f_1(\hat{t})^2}\right) \left(3 \hat{t}^2+\hat{u}^2\right)}{\hat{t}}+\frac{4 \hat{t} \left(f_1(\hat{u})^2+\frac{1}{f_1(\hat{u})^2}\right) \left(\hat{t}^2+3 \hat{u}^2\right)}{\hat{u}}\right. \\ \left.-\frac{2 \left(f_1(\hat{t})+\frac{1}{f_1(\hat{t})}\right) \left(\hat{t}^3-26 \hat{t}^2 \hat{u}+\hat{t} \hat{u}^2-8 \hat{u}^3\right)}{\hat{t}}+\frac{2 \left(f_1(\hat{u})+\frac{1}{f_1(\hat{u})}\right) \left(8 \hat{t}^3-\hat{t}^2 \hat{u}+26 \hat{t} \hat{u}^2-\hat{u}^3\right)}{\hat{u}}\right. \\ \left.-(\hat{t}-\hat{u})^2 \left(f_1(\hat{t}) f_1(\hat{u})+\frac{f_1(\hat{t})}{f_1(\hat{u})}+\frac{1}{f_1(\hat{t}) f_1(\hat{u})}+\frac{f_1(\hat{u})}{f_1(\hat{t})}\right)\right. \\ \left.+\frac{4 \left(6 \hat{t}^4-\hat{t}^3 \hat{u}+38 \hat{t}^2 \hat{u}^2-\hat{t} \hat{u}^3+6 \hat{u}^4\right)}{\hat{t} \hat{u}}\right] \,\,\, .
\end{split}
\end{equation}

\subsection{$gg \rightarrow gg$ scattering cross section}\label{sec:appendixgg}

The scattering cross section $\sigma_{gg\rightarrow gg}$ is complicated, but can be summarized via the following decomposition:
\begin{equation}
\hat{\sigma}_{gg\rightarrow gg} = 
\frac{\alpha_s^2}{\hat{s}} 
\sum_{i,j,k,\ell,m=0}^4 f_2(0,0)^i f_2(\hat{t},0)^j f_2(\hat{u},0)^k f_3(0,\hat{t},0)^\ell f_3(0,\hat{u},0)^m F_{ijk\ell m}(\hat{s},\hat{t},\hat{u})~.
\end{equation}
We find that 
\begin{equation}
F_{ikjm\ell} (\hat{s},\hat{t},\hat{u}) = F_{ijk\ell m}(\hat{s},\hat{u},\hat{t})~,
\end{equation}
that is, there are non-vanishing functions $F$ in addition to those shown below that are obtained by swapping both $j$ and $k$  and $\ell$ and $m$, and whose value is obtained from the result shown by swapping $\hat{t}\leftrightarrow\hat{u}$.   All other other $F_{ijk\ell m}(\hat{s},\hat{t},\hat{u})$ are zero.  We find:

\begin{equation}
F_{00020}(\hat{s},\hat{t},\hat{u}) = \frac{9\hat{t}^2\hat{u}^2(3\hat{t}^2+10\hat{t}\hat{u}+10\hat{u}^2)}{4\hat{s}^2} \,\,\, ,
\end{equation}

\begin{equation}
\begin{split}
F_{40000} (\hat{s},\hat{t},\hat{u}) = \frac{9}{256}\left[ \frac{2\hat{s}^2}{f_1(\hat{s})^2}(\hat{t}-\hat{u})^2+\frac{2\hat{t}^2}{f_1(\hat{t})^2}(\hat{t}+2\hat{u})^2 + \frac{2\hat{u}^2}{f_1(\hat{u})^2}(2\hat{t}+\hat{u})^2\right. \\ \left. -\frac{\hat{s}\hat{t}}{f_1(\hat{s})f_1(\hat{t})}(\hat{t}-\hat{u})(\hat{t}+2\hat{u})+\frac{\hat{s}\hat{u}}{f_1(\hat{s})f_1(\hat{u})}(\hat{t}-\hat{u})(2\hat{t}+\hat{u})\right. \\ \left. +\frac{\hat{t}\hat{u}}{f_1(\hat{t})f_1(\hat{u})}(2\hat{t}+\hat{u})(\hat{t}+2\hat{u}) \right] \,\,\, ,
\end{split}
\end{equation}

\begin{equation}
F_{04000} (\hat{s},\hat{t},\hat{u}) = \frac{9\hat{t}^2\hat{u}^2(5\hat{t}^2+16\hat{t}\hat{u}+16\hat{u}^2)}{8\hat{s}^2f_1(\hat{t})^2} \,\,\, ,
\end{equation}

\begin{equation}
F_{13000}(\hat{s},\hat{t},\hat{u}) = -\frac{9\hat{t}^2\hat{u}^2(\hat{t}+2\hat{u})}{\hat{s}f_1(\hat{t})^2} \,\,\, ,
\end{equation}

\begin{equation}
F_{03000}(\hat{s},\hat{t},\hat{u}) = -\frac{9\hat{t}\hat{u}^2(\hat{t}^2+4\hat{t}\hat{u}+8\hat{u}^2)}{2\hat{s}^2f_1(\hat{t})^2} \,\,\, ,
\end{equation}

\begin{equation}
F_{11010} (\hat{s},\hat{t},\hat{u}) = \frac{9\hat{t}^2\hat{u}^2(\hat{t}+2\hat{u})}{\hat{s}f_1(\hat{t})} \,\,\, ,
\end{equation}

\begin{equation}
F_{10110} (\hat{s},\hat{t},\hat{u}) = \frac{9\hat{t}^2\hat{u}^2(2\hat{t}+5\hat{u})}{4\hat{s}f_1(\hat{u})} \,\,\, ,
\end{equation}

\begin{equation}
\begin{split}
F_{10010} (\hat{s},\hat{t},\hat{u}) = -\frac{9\hat{t}\hat{u}}{8\hat{s}^2}\left[ \frac{-2\hat{s}}{f_1(\hat{t})}(\hat{t}-2\hat{u})(\hat{t}+2\hat{u}) +\frac{\hat{s}}{f_1(\hat{u})}(2\hat{t}^2+3\hat{t}\hat{u}-3\hat{u}^2) \right. \\ \left.+\frac{\hat{s}}{f_1(\hat{s})}(\hat{t}-\hat{u})(2\hat{t}+3\hat{u}) +2(3\hat{t}^3+6\hat{t}^2\hat{u}+4\hat{t}\hat{u}^2+9\hat{u}^3) \right] \,\,\, ,
\end{split}
\end{equation}

\begin{equation}
F_{01010} (\hat{s},\hat{t},\hat{u}) = -\frac{9 \hat{t} \hat{u}^2 \left(\hat{t} f_1(\hat{t}) (\hat{t}+2 \hat{u})-\hat{t}^2-4 \hat{t} \hat{u}-8 \hat{u}^2\right)}{2 \hat{s}^2 f_1(\hat{t})} \,\,\, ,
\end{equation}

\begin{equation}
F_{01001} (\hat{s},\hat{t},\hat{u}) = \frac{9 \hat{t} \hat{u}^2 \left(\hat{s} \hat{t} f_1(\hat{t})-\hat{t}^2+7 \hat{t} \hat{u}+4 \hat{u}^2\right)}{4 \hat{s}^2 f_1(\hat{t})} \,\,\, ,
\end{equation}

\begin{equation}
F_{20010} (\hat{s},\hat{t},\hat{u}) = \frac{9\hat{t}\hat{u}}{32}\left( \frac{2\hat{t}(\hat{t}+2\hat{u})^2}{\hat{s}f_1(\hat{t})} + \frac{\hat{u}(2\hat{t}+\hat{u})(\hat{t}+3\hat{u})}{\hat{s}f_1(\hat{u})} -\frac{(\hat{t}-\hat{u})(2\hat{t}+3\hat{u})}{f_1(\hat{s})} \right) \,\,\, ,
\end{equation}

\begin{equation}
F_{02010} (\hat{s},\hat{t},\hat{u}) = -\frac{9 \hat{t}^2 \hat{u}^2 \left(5 \hat{t}^2+16 \hat{t} \hat{u}+16 \hat{u}^2\right)}{4 \hat{s}^2 f_1(\hat{t})} \,\,\, ,
\end{equation}

\begin{equation}
F_{02001} (\hat{s},\hat{t},\hat{u}) = -\frac{9 \hat{t}^2 \hat{u}^2 \left(9 \hat{t}^2+21 \hat{t} \hat{u}+8 \hat{u}^2\right)}{8 \hat{s}^2 f_1(\hat{t})} \,\,\, ,
\end{equation}

\begin{equation}
F_{00011} (\hat{s},\hat{t},\hat{u}) = \frac{9 \hat{t}^2 \hat{u}^2 \left(5 \hat{t}^2+12 \hat{t} \hat{u}+5 \hat{u}^2\right)}{4 \hat{s}^2} \,\,\, ,
\end{equation}

\begin{equation}
F_{31000} (\hat{s},\hat{t},\hat{u}) = \frac{9\hat{t}\hat{u}}{32f_1(\hat{t})} \left[ -\frac{\hat{s}}{f_1(\hat{s})}(\hat{t}-\hat{u}) +\frac{4\hat{t}}{f_1(\hat{t})}(\hat{t}+2\hat{u}) +\frac{\hat{u}}{f_1(\hat{u})}(2\hat{t}+\hat{u})  \right] \,\,\, ,
\end{equation}

\begin{equation}
\begin{split}
F_{30000} (\hat{s},\hat{t},\hat{u}) = -\frac{9}{64}\left[ \frac{1}{\hat{s}} \left( \frac{\hat{t}}{f_1(\hat{t})}(\hat{t}+2\hat{u})(3\hat{t}^2+\hat{t}\hat{u}+6\hat{u}^2) \right.\right. \\ \left.\left.+\frac{\hat{u}}{f_1(\hat{u})} (2\hat{t}+\hat{u})(6\hat{t}^2+\hat{t}\hat{u}+3\hat{u}^2) \right) + \frac{(\hat{t}-\hat{u})}{f_1(\hat{s})}\left( \frac{\hat{t}}{f_1(\hat{t})} (\hat{t}+2\hat{u})\right.\right. \\ \left.\left. - \frac{\hat{u}}{f_1(\hat{u})}(2\hat{t}+\hat{u}) \right) +\frac{3\hat{s}\hat{t}\hat{u}}{f_1(\hat{t})f_1(\hat{u})}-\frac{4\hat{t}^2}{f_1(\hat{t})^2}(\hat{t}+2\hat{u})-\frac{4\hat{u}^2}{f_1(\hat{u})^2}(2\hat{t}+\hat{u}) \right. \\ \left.- \frac{\hat{s}(\hat{t}-\hat{u})}{f_1(\hat{s})}\left( (\hat{t}-\hat{u})\left(3+\frac{4}{f_1(\hat{s})}\right)-\frac{\hat{t}}{f_1(\hat{t})} +\frac{\hat{u}}{f_1(\hat{u})} \right)  \right] \,\,\, ,
\end{split}
\end{equation}

\begin{equation}
\begin{split}
F_{22000} (\hat{s},\hat{t},\hat{u}) = \frac{9\hat{t}\hat{u}}{64f_1(\hat{t})} \left[ \frac{1}{f_1(\hat{s})}(\hat{t}-\hat{u})(3\hat{t}+4\hat{u}) -\frac{4\hat{t}}{\hat{s}f_1(\hat{t})}(\hat{t}^2+16\hat{t}\hat{u}+16\hat{u}^2) \right. \\ \left. -\frac{\hat{u}}{\hat{s}f_1(\hat{u})}(2\hat{t}+\hat{u})(\hat{t}+4\hat{u}) \right] \,\,\, ,
\end{split}
\end{equation}

\begin{equation}
F_{02200} (\hat{s},\hat{t},\hat{u}) = \frac{9\hat{t}^2\hat{u}^2(16\hat{t}^2+41\hat{t}\hat{u}+16\hat{u}^2)}{16\hat{s}^2f_1(\hat{t})f_1(\hat{u})} \,\,\, ,
\end{equation}

\begin{equation}
\begin{split}
F_{00010} (\hat{s},\hat{t},\hat{u}) = -\frac{9}{8\hat{s}^2}\left[ \frac{\hat{t}\hat{u}}{f_1(\hat{s})} (\hat{t}-\hat{u})(2\hat{t}+3\hat{u}) +\frac{2\hat{u}}{f_1(\hat{t})}(\hat{t}^3+2\hat{t}^2\hat{u}+4\hat{u}^3) \right. \\ \left. +\frac{\hat{t}}{f_1(\hat{u})}(2\hat{t}^3+2\hat{t}^2\hat{u}-\hat{t}\hat{u}^2+3\hat{u}^3) -\hat{t}\hat{u}^2(\hat{t}-\hat{u})f_1(\hat{s}) -2\hat{t}^2\hat{u}(\hat{t}+2\hat{u})f_1(\hat{t}) \right. \\ \left. +\hat{s}\hat{t}\hat{u}^2f_1(\hat{u}) + 2\hat{t}\hat{u}(\hat{t}+2\hat{u})(3\hat{t}+4\hat{u}) \right] \,\,\, ,
\end{split}
\end{equation}

\begin{equation}
F_{21100} (\hat{s},\hat{t},\hat{u}) = \frac{9\hat{t}^2\hat{u}^2}{2f_1(\hat{t})f_1(\hat{u})} \,\,\, ,
\end{equation}

\begin{equation}
\begin{split}
F_{21000} (\hat{s},\hat{t},\hat{u}) = \frac{9\hat{u}}{32}\left[ \frac{8\hat{t}^3+2\hat{t}^2\hat{u}-\hat{t}\hat{u}^2+2\hat{u}^3}{\hat{s}f_1(\hat{t})f_1(\hat{u})} - \frac{(\hat{t}-\hat{u})(5\hat{t}+2\hat{u})}{f_1(\hat{s})f_1(\hat{t})} + \frac{\hat{t}(\hat{t}-\hat{u})}{f_1(\hat{s})} \right. \\ \left. -\frac{4\hat{t}(3\hat{t}^2-5\hat{t}\hat{u}+6\hat{u}^2)}{\hat{s}f_1(\hat{t})} -\frac{4\hat{t}(5\hat{t}^2+2\hat{t}\hat{u}-4\hat{u}^2)}{\hat{s}f_1(\hat{t})^2} -\frac{\hat{t}\hat{u}(2\hat{t}+\hat{u})}{\hat{s}f_1(\hat{u})} \right] \,\,\, ,
\end{split}
\end{equation}

\begin{equation}
F_{11200} (\hat{s},\hat{t},\hat{u}) = -\frac{9\hat{t}^2\hat{u}^2(8\hat{t}+3\hat{u})}{8\hat{s}f_1(\hat{t})f_1(\hat{u})} \,\,\, ,
\end{equation}

\begin{equation}
\begin{split}
F_{20000} (\hat{s},\hat{t},\hat{u}) = \frac{9}{64}\left[ \frac{\hat{t}\hat{u}}{\hat{s}}\left( (\hat{t}+2\hat{u})\frac{f_1(\hat{u})}{f_1(\hat{t})} + (2\hat{t}+\hat{u})\frac{f_1(\hat{t})}{f_1(\hat{u})} \right) +\frac{4(\hat{t}-\hat{u})^2}{f_1(\hat{s})}\left( \frac{3}{f_1(\hat{s})}+2 \right) \right. \\ \left. + \frac{\hat{t}-\hat{u}}{f_1(\hat{s})}\left( \hat{u}f_1(\hat{u}) - \hat{t}f_1(\hat{t})+\frac{2\hat{t}^3-\hat{t}^2\hat{u}-4\hat{t}\hat{u}^2-2\hat{u}^3}{\hat{s}\hat{t}f_1(\hat{t})}+\frac{2\hat{t}^3+4\hat{t}^2\hat{u}+\hat{t}\hat{u}^2-2\hat{u}^3}{\hat{s}\hat{u}f_1(\hat{u})} \right) \right. \\ \left. + \frac{(\hat{t}-\hat{u})f_1(\hat{s})}{\hat{s}}\left( \frac{\hat{t}(\hat{t}+2\hat{u})}{f_1(\hat{t})} - \frac{\hat{u}(2\hat{t}+\hat{u})}{f_1(\hat{u})} \right) + \frac{2\hat{t}^4+2\hat{t}^3\hat{u}+21\hat{t}^2\hat{u}^2+2\hat{t}\hat{u}^3+2\hat{u}^4}{\hat{t}\hat{u}f_1(\hat{t})f_1(\hat{u})} \right. \\ \left. -\frac{4}{\hat{s}}\left( \frac{3\hat{t}^3+4\hat{t}^2\hat{u}+4\hat{t}\hat{u}^2+4\hat{u}^3}{f_1(\hat{t})^2} + \frac{4\hat{t}^3+4\hat{t}^2\hat{u}+4\hat{t}\hat{u}^2+3\hat{u}^3}{f_1(\hat{u})^2} + \frac{\hat{t}(3\hat{t}^2+\hat{t}\hat{u}+18\hat{u}^2)}{f_1(\hat{t})} \right.\right. \\ \left.\left. + \frac{\hat{u}(18\hat{t}^2+\hat{t}\hat{u}+3\hat{u}^2)}{f_1(\hat{u})} \right) + \frac{4}{\hat{s}^2}\left( 15\hat{t}^4+21\hat{t}^3\hat{u}+44\hat{t}^2\hat{u}^2+21\hat{t}\hat{u}^3+15\hat{u}^4 \right) \right] \,\,\, ,
\end{split}
\end{equation}

\begin{equation}
\begin{split}
F_{12000} (\hat{s},\hat{t},\hat{u}) = \frac{9\hat{t}\hat{u}}{16\hat{s}^2f_1(\hat{t})} \left[ \frac{\hat{s}}{f_1(\hat{s})} (\hat{t}-\hat{u})(3\hat{t}+4\hat{u}) - \frac{4\hat{s}}{f_1(\hat{t})} (\hat{t}^2+4\hat{t}\hat{u}-8\hat{u}^2) \right. \\ \left.+ \frac{s}{f_1(\hat{u})} (4\hat{t}^2+7\hat{t}\hat{u}-4\hat{u}^2) + 9\hat{t}^3+13\hat{t}^2\hat{u}+24\hat{u}^3\right] \,\,\, ,
\end{split}
\end{equation}

\begin{equation}
F_{11100} (\hat{s},\hat{t},\hat{u}) = \frac{9\hat{t}\hat{u}}{4\hat{s}}\left[ \frac{2}{f_1(\hat{t})f_1(\hat{u})}(2\hat{t}^2-\hat{t}\hat{u}+2\hat{u}^2)-\hat{t}\hat{u}\left( \frac{1}{f_1(\hat{t})}+\frac{1}{f_1(\hat{u})} \right) \right] \,\,\, ,
\end{equation}

\begin{equation}
F_{02100} (\hat{s},\hat{t},\hat{u}) = \frac{9\hat{t}^3\hat{u}(\hat{u}f_1(\hat{u})-10\hat{t}-19\hat{u})}{8\hat{s}^2f_1(\hat{t})f_1(\hat{u})} \,\,\, ,
\end{equation}

\begin{equation}
\begin{split}
F_{11000} (\hat{s},\hat{t},\hat{u}) = \frac{9}{8\hat{s}^2} \left[ \frac{2\hat{s}^2\hat{u}(\hat{t}-\hat{u})}{f_1(\hat{s})f_1(\hat{t})} - \frac{2\hat{s}(\hat{t}^3-\hat{t}^2\hat{u}+3\hat{t}\hat{u}^2-\hat{u}^3)}{f_1(\hat{t})f_1(\hat{u})} + \frac{\hat{s}\hat{t}\hat{u}(\hat{t}-\hat{u})}{f_1(\hat{s})} \right. \\ \left. - \frac{4\hat{s}\hat{u}(2\hat{t}^2-\hat{t}\hat{u}+2\hat{u}^2)}{f_1(\hat{t})^2} -\frac{\hat{u}(\hat{t}^3+11\hat{t}^2\hat{u}-6\hat{t}\hat{u}^2+12\hat{u}^3)}{f_1(\hat{t})} - \frac{\hat{s}\hat{t}\hat{u}^2}{f_1(\hat{u})} \right. \\ \left. + \frac{\hat{s}\hat{t}\hat{u}}{f_1(\hat{t})}\left( (\hat{t}-\hat{u})f_1(\hat{s})+\hat{u}f_1(\hat{u}) \right) + \hat{t}\hat{u}(3\hat{t}^2+5\hat{t}\hat{u}+6\hat{u}^2) \right] \,\,\, ,
\end{split}
\end{equation}

\begin{equation}
\begin{split}
F_{01100} (\hat{s},\hat{t},\hat{u}) = \frac{9\hat{t}\hat{u}}{4\hat{s}^2}\left[ \frac{1}{f_1(\hat{t})f_1(\hat{u})}(2\hat{t}^2+13\hat{t}\hat{u}+2\hat{u}^2)+\frac{\hat{t}}{f_1(\hat{u})}(2\hat{t}+\hat{u})\right. \\ \left.+\frac{\hat{u}}{f_1(\hat{t})}(\hat{t}+2\hat{u})+\hat{t}\hat{u} \right] \,\,\, ,
\end{split}
\end{equation}

\begin{equation}
\begin{split}
F_{02000} (\hat{s},\hat{t},\hat{u}) = \frac{9}{16\hat{s}^2} \left[ \frac{\hat{t}\hat{u}(\hat{t}-\hat{u})(3\hat{t}+4\hat{u})}{f_1(\hat{s})f_1(\hat{t})} + \frac{\hat{t}(6\hat{t}^3+8\hat{t}^2\hat{u}-3\hat{t}\hat{u}^2+4\hat{u}^3)}{f_1(\hat{t})f_1(\hat{u})} \right. \\ \left. - \frac{\hat{t}^2\hat{u}^2f_1(\hat{u})}{f_1(\hat{t})} + \frac{4\hat{u}(\hat{t}^3+8\hat{t}^2\hat{u}+8\hat{t}\hat{u}^2+16\hat{u}^3)}{f_1(\hat{t})^2} + \frac{\hat{t}^2\hat{u}(\hat{t}-\hat{u})f_1(\hat{s})}{f_1(\hat{t})} \right. \\ \left. + \frac{4\hat{t}\hat{u}(\hat{t}+2\hat{u})(3\hat{t}+4\hat{u})}{f_1(\hat{t})} + 8\hat{t}^2\hat{u}^2 \right]\,\,\, ,
\end{split}
\end{equation}

\begin{equation}
\begin{split}
F_{10000} (\hat{s},\hat{t},\hat{u}) = -\frac{9}{16\hat{s}^2} \left[ \hat{s}(\hat{t}-\hat{u})^2\left( 3f_1(\hat{s})+\frac{1}{f_1(\hat{s})} \right) + \hat{t}f_1(\hat{t})(3\hat{t}^2+5\hat{t}\hat{u}+6\hat{u}^2) \right. \\ \left. +  \hat{u}f_1(\hat{u})(6\hat{t}^2+5\hat{t}\hat{u}+3\hat{u}^2) + 12\hat{s}(\hat{t}^2+\hat{t}\hat{u}+\hat{u}^2) + \frac{2 \hat{s}^2(\hat{t}-\hat{u})}{f_1(\hat{s})} \left( \frac{\hat{u}}{\hat{t}f_1(\hat{t})}-\frac{\hat{t}}{\hat{u}f_1(\hat{u})} \right) \right. \\ \left. -4\hat{s}\left( \frac{(\hat{t}-\hat{u})^2}{f_1(\hat{s})^2} - \frac{\hat{t}^2+2\hat{u}^2}{f_1(\hat{t})^2} - \frac{2\hat{t}^2+\hat{u}^2}{f_1(\hat{u})^2} \right) -\hat{s}(\hat{t}-\hat{u})\left( \frac{\hat{t}f_1(\hat{s})}{f_1(\hat{t})} - \frac{\hat{t}f_1(\hat{t})}{f_1(\hat{s})} - \frac{\hat{u}f_1(\hat{s})}{f_1(\hat{u})} \right. \right. \\ \left. \left.+ \frac{\hat{u}f_1(\hat{u})}{f_1(\hat{s})} \right) - \hat{s}\hat{t}\hat{u}\left( \frac{f_1(\hat{t})}{f_1(\hat{u})} + \frac{f_1(\hat{u})}{f_1(\hat{t})} \right) +\frac{2\hat{s}(\hat{t}^4+5\hat{t}^2\hat{u}^2+\hat{u}^4)}{\hat{t}\hat{u}f_1(\hat{t})f_1(\hat{u})} - \frac{1}{\hat{t}f_1(\hat{t})}(3\hat{t}^4+3\hat{t}^3\hat{u} \right. \\ \left. +24\hat{t}^2\hat{u}^2+8\hat{t}\hat{u}^3+12\hat{u}^4) - \frac{1}{\hat{u}f_1(\hat{u})} (12\hat{t}^4+8\hat{t}^3\hat{u}+24\hat{t}^2\hat{u}^2+3\hat{t}\hat{u}^3+3\hat{u}^4) \right]\,\,\, ,
\end{split}
\end{equation}

\begin{equation}
\begin{split}
F_{01000} (\hat{s},\hat{t},\hat{u}) = -\frac{9}{8\hat{s}^2}\left[ \hat{t}\hat{u}(\hat{t}-\hat{u})\left( f_1(\hat{s})-\frac{1}{f_1(\hat{s})} \right) +\hat{t}\hat{u}^2\left( f_1(\hat{u})-\frac{1}{f_1(\hat{u})} \right) \right. \\ \left. +4\hat{t}^2\hat{u}f_1(\hat{t})+\frac{2\hat{t}^3}{f_1(\hat{u})}+\frac{\hat{u}}{f_1(\hat{s})f_1(\hat{t})} (\hat{t}-\hat{u})(\hat{t}+2\hat{u}) \right. \\ \left.+\frac{1}{f_1(\hat{t})f_1(\hat{u})}(2\hat{t}^3+8\hat{t}^2\hat{u}-\hat{t}\hat{u}^2+2\hat{u}^3) - \frac{4\hat{u}}{\hat{t}f_1(\hat{t})^2}(\hat{t}^3-2\hat{t}^2\hat{u}-2\hat{t}\hat{u}^2-4\hat{u}^3) \right. \\ \left.+\frac{\hat{u}}{f_1(\hat{t})}\left( f_1(\hat{s})(\hat{t}-\hat{u})(3\hat{t}+2\hat{u})+\hat{u}f_1(\hat{u})(\hat{t}+2\hat{u})+4(\hat{t}+2\hat{u})(2\hat{t}+3\hat{u}) \right) \right]\,\,\, ,
\end{split}
\end{equation}

\begin{equation}
\begin{split}
F_{00000} (\hat{s},\hat{t},\hat{u}) = -\frac{9}{16\hat{s}^2} \left[ \frac{(\hat{t}-\hat{u})}{f_1(\hat{s})}\left( \hat{t}f_1(\hat{t})-\hat{u}f_1(\hat{u})+4(\hat{t}-\hat{u}) - \frac{2(\hat{t}-\hat{u})}{f_1(\hat{s})} \right. \right. \\ \left. \left.  -  \frac{\hat{t}^2+2\hat{t}\hat{u}+\hat{u}^2}{\hat{t}f_1(\hat{t})} + \frac{2\hat{t}^2+2\hat{t}\hat{u}+\hat{u}^2}{\hat{u}f_1(\hat{u})} \right) - 20\left( \hat{t}^2+\hat{t}\hat{u}+\hat{u}^2 \right) \right. \\ \left. +\frac{1}{\hat{t}f_1(\hat{t})}\left[ (\hat{t}-\hat{u})(\hat{t}^2-2\hat{t}\hat{u}-2\hat{u}^2)f_1(\hat{s}) +\hat{u}(\hat{t}^2-2\hat{u}^2)f_1(\hat{u})-4\hat{u}(\hat{t}+2\hat{u})^2 \right] \right. \\ \left. + \frac{1}{\hat{u}f_1(\hat{u})}\left[ (\hat{t}-\hat{u})(2\hat{t}^2+2\hat{t}\hat{u}-\hat{u}^2)f_1(\hat{s}) - \hat{t}(2\hat{t}^2-\hat{u}^2)f_1(\hat{t}) -4\hat{t}(2\hat{t}+\hat{u})^2 \right] \right. \\ \left. -\frac{2}{\hat{t}^2f_1(\hat{t})^2} \left( \hat{t}^4+4\hat{t}^2\hat{u}^2+4\hat{t}\hat{u}^3+4\hat{u}^4 \right) - \frac{2}{\hat{u}^2f_1(\hat{u})^2}\left( 4\hat{t}^4+4\hat{t}^3\hat{u}+4\hat{t}^2\hat{u}^2+\hat{u}^4 \right) \right. \\ \left. -(\hat{t}-\hat{u})f_1(\hat{s})\left( (\hat{t}-\hat{u})(2f_1(\hat{s})-4) + \hat{t}f_1(\hat{t})-\hat{u}f_1(\hat{u}) \right) -\hat{t}\hat{u}f_1(\hat{t})f_1(\hat{u}) \right. \\ \left. -2\left( \hat{t}^2f_1(\hat{t})^2 + \hat{u}^2f_1(\hat{u})^2 \right) - \frac{1}{\hat{t}\hat{u}f_1(\hat{t})f_1(\hat{u})} \left( 2\hat{t}^4+4\hat{t}^3\hat{u} +13\hat{t}^2\hat{u}^2+4\hat{t}\hat{u}^3+2\hat{u}^4 \right) \right]\,\,\, .
\end{split}
\end{equation}


\begin{thebibliography}{99}

\bibitem{Grinstein:2007mp}
B.~Grinstein, D.~O'Connell and M.~B.~Wise,
``The Lee-Wick standard model,''
Phys.\ Rev.\ D \textbf{77}, no.~2, 025012 (2008),
\href{http://arxiv.org/abs/0704.1845}{arXiv:0704.1845 [hep-ph]}.

\bibitem{LeeWick:1969}
T.~D.~Lee and G.~C.~Wick,
``Negative  metric and the unitarity of the S-matrix,'' Nucl.\ Phys.\ B \textbf{9} (1969), 209;
``Finite theory of quantum electrodynamics,'' Phys.\ Rev.\ D \textbf{2}, no.~6, 1033 (1970).

\bibitem{Cutkosky:1969fq}
R.~E.~Cutkosky, P.~V.~Landshoff, D.~I.~Olive and J.~C.~Polkinghorne,
``A non-analytic S-matrix,''
Nucl.\ Phys.\ B \textbf{12}, 281--300 (1969).

\bibitem{Anselmi}
D.~Anselmi and M.~Piva,
``A new formulation of Lee-Wick quantum field theory,''
JHEP \textbf{06}, 066 (2017),
\href{https://arxiv.org/abs/1703.04584}{arXiv:1703.04584 [hep-th]};
D.~Anselmi and M.~Piva,
``Perturbative unitarity of Lee-Wick quantum field theory,''
Phys.\ Rev.\ D \textbf{96}, 045009 (2017),
\href{https://arxiv.org/abs/1703.05563}{arXiv:1703.05563 [hep-th]}.

\bibitem{Efimov:1967}
G.~V.~Efimov,
``Nonlocal quantum theory of the scalar field,''
Commun.\ Math.\ Phys.\ \textbf{5}, 42--56 (1967).

\bibitem{Krasnikov:1987}
N.~V.~Krasnikov,
``Nonlocal gauge theories,''
Theor.\ Math.\ Phys.\ \textbf{73}, 1184--1190 (1987).

\bibitem{Kuzmin:1989}
Yu.~V.~Kuz'min,
``Convergent nonlocal gravitation,''
Sov.\ J.\ Nucl.\ Phys.\ \textbf{50}, no.~6, 1011--1014 (1989).

\bibitem{Tomboulis:1997gg}
E.~T.~Tomboulis,
``Super-renormalizable gauge and gravitational theories,''
\href{http://arxiv.org/abs/hep-th/9702146}{arXiv:hep-th/9702146 [hep-th]}.

\bibitem{Modesto:2011kw}
L.~Modesto,
``Super-renormalizable Quantum Gravity,''
Phys.\ Rev.\ D \textbf{86}, no.~4, 044005 (2012),
\href{http://arxiv.org/abs/1107.2403}{arXiv:1107.2403 [hep-th]}.

\bibitem{Biswas:2011ar}
T.~Biswas, E.~Gerwick, T.~Koivisto and A.~Mazumdar,
``Towards singularity and ghost free theories of gravity,''
Phys.\ Rev.\ Lett.\ \textbf{108}, no.~3, 031101 (2012),
\href{http://arxiv.org/abs/1110.5249}{arXiv:1110.5249 [gr-qc]}.

\bibitem{Buoninfante:2018mre}
L.~Buoninfante, G.~Lambiase and A.~Mazumdar,
``Ghost-free infinite-derivative quantum field theory,''
Nucl.\ Phys.\ B \textbf{944}, 114646 (2019),
\href{http://arxiv.org/abs/1805.03559}{arXiv:1805.03559 [hep-th]}.

\bibitem{Boos:2021chb}
J.~Boos and C.~D.~Carone,
``Asymptotic nonlocality,''
Phys.\ Rev.\ D \textbf{104}, 015028 (2021),
\href{http://arxiv.org/abs/2104.11195}{arXiv:2104.11195 [hep-th]}.

\bibitem{Boos:2021jih}
J.~Boos and C.~D.~Carone,
``Asymptotic nonlocality in gauge theories,''
Phys.\ Rev.\ D \textbf{104}, 095020 (2021),
\href{http://arxiv.org/abs/2109.06261}{arXiv:2109.06261 [hep-th]}.

\bibitem{Boos:2021lsj} 
J.~Boos and C.~D.~Carone,
``Asymptotic nonlocality in non-Abelian gauge theories,''
Phys.\ Rev.\ D \textbf{105}, 035034 (2022),
\href{https://arxiv.org/abs/2112.05270}{arXiv:2112.05270 [hep-ph]}.

\bibitem{Boos:2022biz} 
J.~Boos and C.~D.~Carone,
``Asymptotically nonlocal gravity,''
JHEP \textbf{06}, 017 (2023),
\href{https://arxiv.org/abs/2212.00861}{arXiv:2212.00861 [hep-th]}.

\bibitem{Carone:2023cnp}
C.~D.~Carone and M.~R.~Musser,
``Note on scattering in asymptotically nonlocal theories,''
Phys. Rev. D \textbf{108}, no.9, 095015 (2023),
\href{https://arxiv.org/abs/2308.11051}{arXiv:2308.11051 [hep-th]}.

\bibitem{Pais:1950za}
A.~Pais and G.~Uhlenbeck,
``On field theories with nonlocalized action,''
Phys.\ Rev.\ \textbf{79}, 145 (1950).

\bibitem{Carone:2008iw}
C.~D.~Carone and R.~F.~Lebed,
``A higher-derivative Lee-Wick standard model,''
JHEP \textbf{01}, 043 (2009),
\href{http://arxiv.org/abs/0811.4150}{arXiv:0811.4150 [hep-ph]}.

\bibitem{NLunitarity}
 F.~Briscese and L.~Modesto,
``Nonunitarity of Minkowskian nonlocal quantum field theories,''
Eur.\ Phys.\ J.\ C \textbf{81}, no.~8, 730 (2021),
\href{http://arxiv.org/abs/2103.00353}{arXiv:2103.00353 [hep-th]};
A.~S.~Koshelev and A.~Tokareva,
``Unitarity of Minkowski nonlocal theories made explicit,''
Phys.\ Rev.\ D \textbf{104}, no.~2, 025016 (2021),
\href{http://arxiv.org/abs/2103.01945}{arXiv:2103.01945 [hep-th]};
F.~Briscese and L.~Modesto,
``Cutkosky rules and perturbative unitarity in Euclidean nonlocal quantum field theories,''
Phys.\ Rev.\ D \textbf{99}, no.~10, 104043 (2019),
\href{http://arxiv.org/abs/1803.08827}{arXiv:1803.08827 [gr-qc]};
R.~Pius and A.~Sen,
``Cutkosky rules for superstring field theory,''
JHEP \textbf{10}, 024 (2016)
[Erratum: JHEP \textbf{09}, 122 (2018)],
\href{http://arxiv.org/abs/1604.01783}{arXiv:1604.01783 [hep-th]};
C.~D.~Carone,
``Unitarity and microscopic acausality in a nonlocal theory,''
Phys.\ Rev.\ D \textbf{95}, no.~4, 045009 (2017),

\bibitem{Biswas:2014yia}
T.~Biswas and N.~Okada,
``Towards LHC physics with nonlocal Standard Model,''
Nucl. Phys. B \textbf{898}, 113-131 (2015),
\href{https://arxiv.org/abs/1407.3331}{arXiv:1407.3331 [hep-ph]}.

\bibitem{Bertram:1998wf}
I.~Bertram and E.~H.~Simmons,
``Dijet mass spectrum limits on flavor universal colorons,''
Phys. Lett. B \textbf{443}, 347-351 (1998),
\href{https://arxiv.org/abs/hep-ph/9809472}{arXiv:hep-ph/9809472 [hep-ph]}.

\bibitem{FeynCalc}
V.~Shtabovenko, R.~Mertig, and F.~Orellana,
``FeynCalc 10: Do multiloop integrals dream of computer codes?",
\href{https://arxiv.org/abs/2312.14089}{arXiv:2312.14089 [hep-ph]}.
V.~Shtabovenko, R.~Mertig, and F.~Orellana,
``FeynCalc 9.3: New features and improvements,"
\href{https://arxiv.org/abs/2001.04407}{arXiv:2001.04407 [hep-ph]}.
V.~Shtabovenko, R.~Mertig, and F.~Orellana,
``New developments in FeynCalc 9.0,"
Comput. Phys. Commun., \textbf{207}, (2016),
\href{https://arxiv.org/abs/1601.01167}{arXiv:1601.01167 [hep-ph]}.
R.~Mertig, M.~Bohm, and A.~Denner,
``Feyn Calc - Computer-algebraic calculation of Feynman amplitudes,"
Comput. Phys. Commun., \textbf{64}, (1991).

\bibitem{schwinn}
See, for example, C.~Schwinn, ``Modern Methods of Quantum Chromodynamics,"  Lecture notes,
Albert-Ludwigs-Universit\"{a}t, Freiburg, 2015, 
\href{https://www.tep.physik.uni-freiburg.de/lectures/archive/QCD-WS-14/qcd}{https://www.tep.physik.uni-freiburg.de/lectures/archive/QCD-WS-14/qcd}.

\bibitem{Eichten:1984eu}
E.~Eichten, I.~Hinchliffe, K.~D.~Lane and C.~Quigg,
``Super Collider Physics,''
Rev. Mod. Phys. \textbf{56}, 579-707 (1984).

\bibitem{Peskin:1995ev}
M.~E.~Peskin and D.~V.~Schroeder,
``An Introduction to quantum field theory,''
Addison-Wesley, 1995,
ISBN 978-0-201-50397-5.

\bibitem{CMS:2019gwf}
A.~M.~Sirunyan \textit{et al.} [CMS],
``Search for high mass dijet resonances with a new background prediction method in proton-proton collisions at $\sqrt{s} =$ 13 TeV,''
JHEP \textbf{05}, 033 (2020),
\href{https://arxiv.org/abs/1911.03947}{arXiv:1911.03947 [hep-ex]}.

\bibitem{CMS:2018mgb}
A.~M.~Sirunyan \textit{et al.} [CMS],
``Search for narrow and broad dijet resonances in proton-proton collisions at $ \sqrt{s}=13 $ TeV and constraints on dark matter mediators and other new particles,''
JHEP \textbf{08}, 130 (2018),
\href{https://arxiv.org/abs/1806.00843}{arXiv:1806.00843 [hep-ex]}.

\bibitem{ATLAS:2018qto}
M.~Aaboud \textit{et al.} [ATLAS],
``Search for low-mass dijet resonances using trigger-level jets with the ATLAS detector in $pp$ collisions at $\sqrt{s}=13$ TeV,''
Phys. Rev. Lett. \textbf{121}, no.~8, 081801 (2018),
\href{https://arxiv.org/abs/1804.03496}{arXiv:1804.03496 [hep-ex]}.

\bibitem{Clark:2016jgm}
D.~B.~Clark, E.~Godat and F.~I.~Olness,
``ManeParse : A Mathematica  reader for Parton Distribution Functions,''
Comput. Phys. Commun. \textbf{216}, 126-137 (2017),
\href{https://arxiv.org/abs/1605.08012}{arXiv:1605.08012 [hep-ph]}.

\bibitem{Kovarik:2015cma}
K.~Kovarik, A.~Kusina, T.~Jezo, D.~B.~Clark, C.~Keppel, F.~Lyonnet, J.~G.~Morfin, F.~I.~Olness, J.~F.~Owens and I.~Schienbein, \textit{et al.}
``nCTEQ15 - Global analysis of nuclear parton distributions with uncertainties in the CTEQ framework,''
Phys. Rev. D \textbf{93}, no.~8, 085037 (2016),
\href{https://arxiv.org/abs/1509.00792}{arXiv:1509.00792 [hep-ph]}.

\bibitem{Frederix:2016ost}
R.~Frederix, S.~Frixione, V.~Hirschi, D.~Pagani, H.~S.~Shao and M.~Zaro,
``The complete NLO corrections to dijet hadroproduction,''
JHEP \textbf{04}, 076 (2017),
\href{https://arxiv.org/abs/1612.06548}{arXiv:1612.06548 [hep-ph]}.

\end{thebibliography}
\end{document}